\documentclass[prl,aps,showpacs,twocolumn]{revtex4}

\usepackage{graphicx}
\usepackage{epsfig}

\begin{document}

\title{Production of Long-Lived Ultracold Li$_{2}$ Molecules from a
Fermi gas} \author{J.\,Cubizolles$^1$, T.\,Bourdel$^1$,
S.\,J.\,J.\,M.\,F.\,Kokkelmans$^1$, G.\,V.\,Shlyapnikov$^{1,2,3}$ and
C.\,Salomon$^1$} \affiliation{{$^1$Laboratoire Kastler Brossel, Ecole
Normale Sup\'erieure, 24 rue Lhomond, 75231 Paris 05, France}\\{$^2$FOM
Institute AMOLF, Kruislaan 407, 1098 SJ Amsterdam, The
Netherlands}\\{$^3$Russian Research Center Kurchatov Institute,
Kurchatov Square, 123182 Moscow, Russia}} \date{\today}

\begin{abstract}
We create weakly-bound Li$_2$ molecules from a degenerate two component
Fermi gas by sweeping a magnetic field across a Feshbach resonance. The
atom-molecule transfer efficiency can reach 85\,$\%$ and is studied as a
function of magnetic field and initial temperature. The bosonic
molecules remain trapped for $0.5\,$s and their temperature is within a
factor of 2 from the Bose-Einstein condensation temperature. A
thermodynamical model reproduces qualitatively the experimental
findings.
\end{abstract}


\maketitle

Feshbach resonances constitute a unique tool to tune the microscopic
interactions in ultracold bosonic and fermionic gases
\cite{Feshbach,Naturespecial,Inouye98}. These resonances arise when the
total energy of a pair of colliding atoms matches the energy of a bound
state of another hyperfine manifold, leading to the resonant occupation
of this state during the collision. Thus, by means of an external
magnetic field, one is able to change the magnitude and sign of the
scattering length. In bosonic samples, the collapse of Bose-Einstein
condensates (BEC) for negative scattering length \cite{Jilacollapse},
soliton formation \cite{Khaykovich02,Truscott02}, and coherent
oscillations between an atomic condensate and molecules \cite{Donley02}
have been observed. For fermions with attractive interaction, the
superfluid transition temperature is predicted to be maximum near a
Feshbach resonance
\cite{Nozieres85,Randeria93,Holland01,Ohashi02,Milstein02}.

In this letter we give another striking example of the control of
interactions in a Fermi gas. We perform time-dependent experiments near
a Feshbach resonance to produce in a reversible manner ultracold and
trapped molecules from a quantum degenerate fermionic $^6$Li gas. The
production efficiency exceeds 80\% and the observed molecule lifetime
reaches half a second. The phase-space density of these bosonic
molecules is on the order of one, the highest value reported so far.
Using a similar method with fermionic $^{40}$K atoms, the JILA group
recently reported molecule production with a lifetime of 1\,ms, atom to
molecule conversion efficiency of 50\% and direct measurement of the
molecular binding energy \cite{Regal03}. Molecules have also recently
been produced from $^{87}$Rb and $^{133}$Cs condensates \cite{Grimm03,
Durr03} and also from a cold $^{133}$Cs cloud \cite{Chin03}. Molecule
formation has also been achieved through one-photon or two-photon
photoassociation but with a considerably lower phase-space density than
reported here \cite{Wynar00}. Our work paves the way to Bose-Einstein
condensation of molecules and to the study of the crossover between the
regime of molecular BEC and the regime of superfluid BCS pairing in
Fermi systems \cite{Nozieres85,Randeria93,Ohashi02,Milstein02}.

\begin{figure}[hb]
\begin{center}
  \epsfig{file=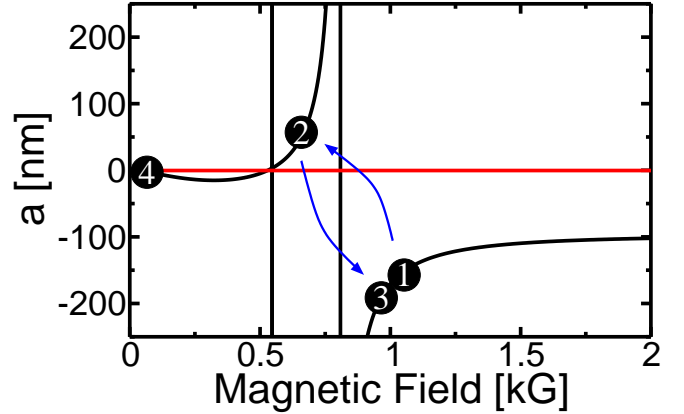, width=\linewidth}
\caption{ \label{fig:figure1} Calculated scattering length $a$ versus
magnetic field for the $^6$Li $|F,m_F\rangle=|1/2,1/2\rangle$,
$|1/2,-1/2\rangle$ mixture near the 810\,Gauss Feshbach
resonance. Scanning over the resonance from 1 to 2 in 50 \,ms produces
cold and trapped molecules with an efficiency up to $85\%$, resulting in
an almost complete disappearance of the atomic signal at 2. Reversing
then the scan to the initial position re-establishes the initial atomic
signal. In all cases, atoms are detected and imaged at position 4 after
abrupt switch-off of the $B$ field.}
\end{center}
\end{figure}

In $^6$Li a broad ($\simeq 100$ Gauss) Feshbach resonance exists between
the two Zeeman sublevels of the hyperfine ground state : $|1/2,-1/2
\rangle$ and $|1/2,1/2 \rangle$ at a magnetic field of 810 Gauss, see
Fig.\,\ref{fig:figure1} \cite{Gupta03,Bourdel03}. In a recent study of
this Feshbach resonance, we have reported an anomalous negative value
for the gas interaction energy between 700 Gauss and 810 gauss,
i.e. below resonance, where the scattering length $a$ is positive
\cite{Bourdel03}. We suggested that this could be due to the presence of
weakly bound molecules confined simultaneously with the cloud of
ultracold fermions, and a recent theoretical paper explains our results
\cite{Kokkelmans03}.

The method used here to produce molecules is illustrated in
Fig.\ref{fig:figure1} and was suggested for Bose gases in
\cite{Timmermans98,Abeelen99, Julienne}. It consists in scanning over a
Feshbach resonance from the region of attractive interaction ($a<0$) to
region\,2 in Fig.\,\ref{fig:figure1} where $a$ is large and positive,
and where a weakly bound molecular state exists with energy
$E_{b}=-\hbar^2/ma^2$ (where $m$ is the atomic mass). It is thus
energetically favorable to populate this bound state. Having twice the
polarisability of individual atoms, the molecules are expected to be
confined in an optical dipole trap with the same oscillations
frequencies as atoms and twice the trap depth.

Experimentally, using the setup described in \cite{Bourdel03}, we
prepare $N_1=N_\uparrow+N_\downarrow=1.5\,10^5$ atoms in a $50$($10
$)$\%$ mixture of $|1/2,-1/2 \rangle$ and $|1/2,1/2 \rangle$ in a
crossed Nd:YAG dipole trap at $B=1060\,G$ where the scattering length
between $|1/2,-1/2 \rangle$ and $|1/2,1/2 \rangle$ is large and negative
$a=-150\,$nm (position 1 in Fig.\,\ref{fig:figure1}). For a power of
$1.53\,$W (resp.  $4.3\,$W) in the horizontal (resp. vertical) trapping
beam propagating along Ox (resp. Oy), oscillation frequencies are
$\omega_{x}/2\pi= 2.9(3)\,$kHz, $\omega_{y}/2\pi=5.9(5)\,$kHz and
$\omega_{z}/2\pi= 6.5(6)\,$kHz. By evaporation in the optical trap, the
temperature $T$ of the gas mixture can be tuned between $0.2$ and
$0.5\,T_{F}$ where $T_{F}$ is the Fermi temperature defined by
$T_{F}=\hbar \bar\omega\,(3 N_1)^{1/3}{}/k_{B}$ and
$\overline{\omega}=(\omega_{x} \omega_{y} \omega_{z})^{1/3}$.

The Fermi gas quantum degeneracy is measured through analysis of
absorption images after abrupt ($20\,\mu$s) switch-off of the magnetic
field and time of flight expansion of the cloud. Free $^6$Li atoms are
detected at zero magnetic field, for which $a\simeq 0$, using laser
light tuned to the D2 line, position 4 in
Fig.\,\ref{fig:figure1}. Having prepared the degenerate Fermi gas at
position 1, we sweep the magnetic field in 50 ms to position 2 where $a$
is large and positive, the region where a weakly-bound molecular state
exists. Absorption images indeed reveal that the number of atoms $N_2$
as counted after $B$ switch-off to position 4 as explained above, has
dramatically decreased to $\simeq$20$\%$ of the initial number
$N_1$. After letting the system at position 2 for a variable wait time
$t_{\rm w}$ between $0$ and 5\,s, we sweep the magnetic field back to
position 1 in 50\,ms and count the number of atoms, $N_3$, after this
round-trip through resonance.

Surprisingly, when position 2 is at $689\, $ Gauss and $t_{\rm w}=0$, we
find $N_3\simeq N_1$, indicating that no loss has occurred in
100\,ms. This proves that the decrease in detected atom number at
position 2 is not due to losses. We are thus led to conclude that: (i)
at position 2, 80$\%$ of the atoms were in a molecular bound state which
is not detected using light resonant with the atoms at position 4, (ii)
the atom-to-molecule formation process is reversible.  $(N_3-N_2)/2$
represents thus the number of molecules at position 2.

In comparison to other recent experiments on bosons and fermions
\cite{Donley02, Durr03, Regal03, Grimm03}, our 50\,ms scan over the
Feshbach resonance is very slow and, in particular, much longer than the
collision time in the cloud. This allows the system to relax towards
thermal equilibrium between atoms and molecules during the magnetic
field ramp.
\begin{figure}[htb]
\begin{center}
  \epsfig{file=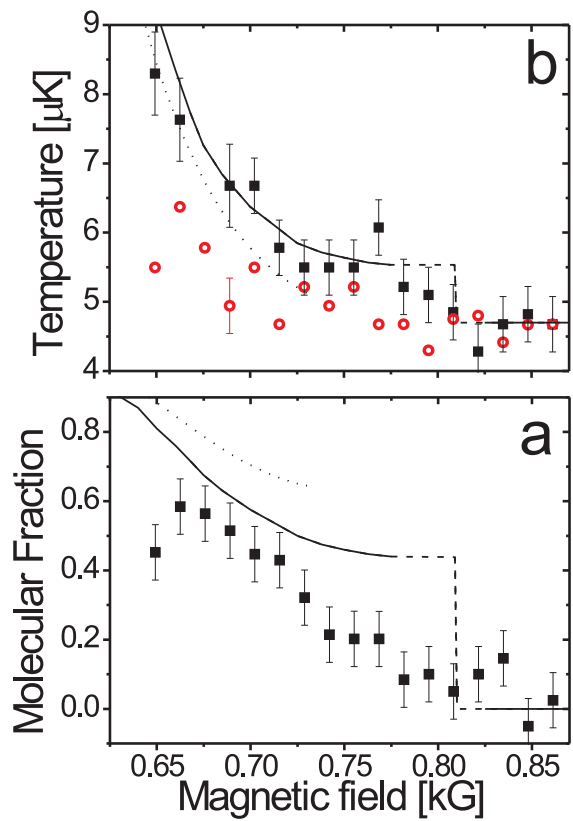, width=.9\linewidth}
\caption{ \label{fig:figure2} a: fraction of atoms in molecular
bound state, $(N_3-N_2)/N_3$,  versus magnetic field at position 2
in Fig.\,1. b: Corresponding atom temperature at position 2
(squares) and 3 (circles). Solid (dotted) line corresponds to the
non-interacting (interacting) thermodynamical model described in text.}
\end{center}
\end{figure}

The relative fraction of atoms bound in a molecule, $(N_3-N_2)/N_3$, and
the temperature of the atoms at position 2 and 3 are plotted in
Fig.\,\ref{fig:figure2} as a function of the magnetic field at position
2. The initial parameters at $1060\,$G for this experiment are
$T=4.7\,\mu$K, $T_F=11\,\mu$K, and $N_1=8\,10^4$. The frequencies of the
trap are $\omega_{x}/2\pi=2.2(2)\,$kHz, $\omega_{y}/2\pi= 4.6(4)\,$kHz,
$\omega_{z}/2\pi= 5.1(5)\,$kHz.  One observes that very few molecules
are detected above $0.77\,$kG. When decreasing the magnetic field, the
fraction of molecules increases up to $60\%$. Below $650\,$G, losses
become important and the molecular fraction decreases, an effect that we
will study below.

A key parameter in our detection scheme is the switch off time
($20\,\mu$s) of the Feshbach magnetic field from position 2. Indeed, the
molecules may or may not dissociate in this process.  Molecules will
dissociate if the relative rate of change of the binding energy
$\text{d}E_b/E_b \text{d}t$ is greater than the oscillation pulsation
$E_b/\hbar$. So, the dissociation of molecules is governed by the
parameter $\alpha=\hbar\,\text{d}E_b/{E_b}^2\text{d}t$. If $\alpha \gg
1$, they dissociate and appear as free atoms in the absorption image.
Our detection scheme is unable to prove molecule formation. On the
contrary, if the binding energy is sufficiently large at position 2
($\alpha \ll 1$), the molecular state follows adiabatically the change
of $B$ to a deeper bound state which is off-resonant with the detection
laser. In this case, $N_3 >N_2$ is the signature of molecule formation.

For $B=700\,$G, one has $\alpha \simeq 1$. Reducing the rate
$\text{d}B/\text{d}t$ by a factor of $\simeq 500$, we observe a shift of
the threshold for molecule detection to $800\,$G, as expected from the
expression of $\alpha$ and the position of the Feshbach resonance at
$810\,$G \cite{Bourdel03}.  As a second consequence of the role of the
switch-off time, we attribute the decrease of the molecular fraction
above $720\,$G in Fig.\,\ref{fig:figure2}a to the crossover between the
adiabatic and non adiabatic regimes.

The temperature of the atoms at position 2 and 3 is plotted in
Fig.\,\ref{fig:figure2}b. A decrease of the magnetic field increases the
binding energy of the molecules. Therefore, the molecular fraction
increases and so does the temperature.  However, sweeping back the
magnetic field to the initial position, the temperature returns close to
its initial value, demonstrating the reversibility of the molecular
formation process. According to \cite{Petrov03}, in our experiment the
rate of 3-body recombination to the weakly bound state is greater than
$10^4\,$Hz (for $B > 665\,$G).  Moreover, the atom-atom, atom-molecule
and molecule-molecule \cite{Petrovprivate} collision rates are also
large compared to the rate at which the resonance is scanned. Therefore
the system remains close to thermal equilibrium.  During the molecule
formation process, the released binding energy must be converted into
kinetic energy, therefore the increase of temperature in
Fig.\,\ref{fig:figure2}b, as $a$ decreases, is expected.

\begin{figure}[htb]
\begin{center}
  \epsfig{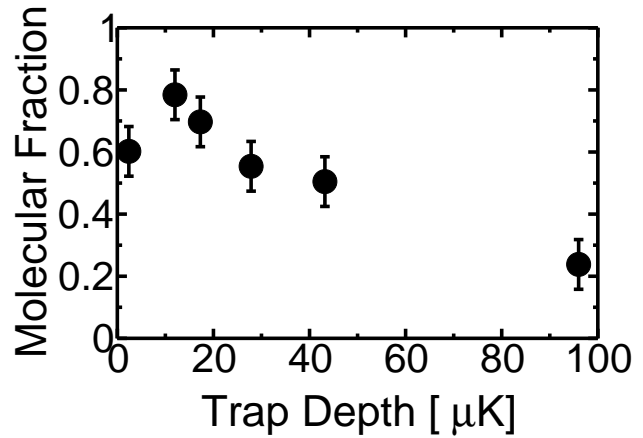}
\caption{ \label{fig:figure3} Fraction of atoms bound in a
molecule as a function of the depth of the dipole trap for
$B=689\,$G, $a= 78\,$ nm, $E_{b}=12\,\mu$K.}
\end{center}
\end{figure}

Interestingly, the efficiency of molecule formation strongly depends on
the trap depth and reaches values up to $80\,\%$ in
Fig.\,\ref{fig:figure3}, the highest efficiency reported so far.  The
trap depth is slowly (800 ms) reduced at $B=1060\,$G, before the
magnetic field sweep through resonance. The gas temperature and the
Fermi temperature decrease by evaporation and adiabatic cooling due to
the reduction of the trap oscillation frequencies.  At $689\,$G,
$a=78\,$nm, $E_{b}/k_{\rm B}=12\,\mu$K. When reducing the trap depth,
$T$ and $T_F$ become small compared to $E_{b}$ and the molecular
fraction increases towards 1. Other experiments show that both
evaporation (reduction of the ratio $T/T_{\rm F}$), and adiabatic
cooling (reduction of $T$ and $T_{\rm F}$ by the same factor)
effectively lead to an increase of the molecular fraction. For instance,
at constant $T_F=11\mu$K, as the temperature is decreased from
$0.5\,T_{\rm F}$ to $0.3\,T_{\rm F}$, we find that the molecular
fraction at $689\,$G increases from $40\,\%$ to $65\,\%$.  Similarly, at
constant $T/T_{\rm F}=0.3$, as $T$ varies from $4.5\,\mu$K to
$1.8\,\mu$K, it increases from $50\,\%$ to $85\,\%$.

Our results can be understood within a thermodynamical model assuming
that atoms and molecules are in equilibrium in a trap during the
magnetic field sweep \cite{Kokkelmans03}. We assume that the sweep is
adiabatic, so the entropy is conserved and it is calculated for the
initial situation where no molecules are present at $B=1060\,$G. The
atom-atom, atom-molecule, molecule-molecule interactions were either
neglected (solid line in Fig.\,\ref{fig:figure2}), or included via a
mean field contribution proportional to $a$ \cite{Petrov03,
Petrovprivate}, in the regime where it is allowed (dotted line). The
molecular fraction is then determined by a set of two parameters
$T/T_{\rm F}$ and $E_b/T_{\rm F}$. As the binding energy gets more
negative, the fraction of molecules increases since the molecular state
becomes energetically more favorable. Moreover, the temperature
increases as each molecule formation event transfers the binding energy
into kinetic energy.  Both models agree fairly well with our data in the
region where both the detection of molecules is efficient and losses are
unimportant ($660\lesssim B\lesssim 720\,$G). In the interacting case, a
perfect agreement is achieved if we allow for a 10$\%$ mismatch in the
quality of the spin mixture.  In the strongly interacting region where
$na^3\geq 1$, the non-interacting model is shown with a dashed line
because we do not expect our model of separated gases of fermionic atoms
and bosonic molecules to be valid. Those models also account
qualitatively for the increase in the molecular fraction with the
evaporation in Fig.\,\ref{fig:figure3}. Indeed, as the trap depth is
lowered, $T$ decreases, the ratio $E_{b}/T$ increases and the bound
state will then have a higher occupancy.

\vspace{0.5cm}
\begin{figure}[ht]
\begin{center}
  \epsfig{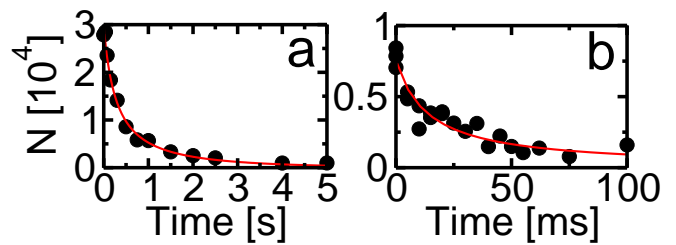}
\caption{ \label{fig:figure4} Decay of Li$_2$ molecules in the
optical trap for two values of the magnetic field: (a) $B=689\,$G,
$a=78\,$nm, (b) $B=636\,$G, $a=35\,$nm. Note the difference
in time scales. Solid lines are fits with 2-body and 1-body
processes giving initial time constants of 500\,ms in(a) and
20\,ms in (b).}
\end{center}
\end{figure}

In order to investigate prospects for the formation of a molecular BEC,
two questions immediately arise: what are the lifetime and degeneracy
parameter of the produced molecular cloud ?  The decay of the molecules
in the optical trap is found by measuring the quantity $(N_3-N_2)/2 $ as
a function of the wait time $t_{\rm w}$ at position 2 before returning
to 3 for the detection process.  Fig.\,\ref{fig:figure4} presents the
number of remaining molecules as a function of the time elapsed at point
2 in a trap with frequencies $\omega_{x}/2\pi= 0.95(10)\,$kHz,
$\omega_{y}/2\pi= 2.0(2)\,$kHz, $\omega_{z}/2\pi= 2.2(2)\,$kHz. In Fig.
\ref{fig:figure4} a and b, the initial temperatures at position 1 are
the same, $1.1\,\mu$K $=0.3\,T_{\rm F}$. Fig \ref{fig:figure4}a
corresponds to $B=689\,G$, $a=78\,$nm, $E_{b}/k_{B}=12\,\mu$K, whereas
Fig.\,\ref{fig:figure4}b corresponds to $B=636\,G$, $a=35\,$nm, and
$E_{b}/k_{B}=60\,\mu$K.  The initial fraction of unpaired atoms (not
engaged in a molecule) is $17\%$ in (a) and $25\%$ in (b).  For a change
in $a$ by the modest factor of $2.2$, the two samples exhibit strikingly
different lifetimes. Decays are well fitted by 2-body + 1-body loss
processes with initial lifetimes of $\simeq 500\,$ms in (a) ($689\,$G)
and $20\,$ms in (b) whereas the 1-body decay exceeds 4 seconds.

A strong decrease in the lifetime of molecules with decreasing $a$ is
expected assuming that they undergo collisional relaxation to deep bound
states \cite{Petrovprivate}. However, the observed factor of 25 in
lifetimes is surprisingly large. At $689\,$G, we can evaluate the 2-body
loss rate to be $G=2.4_{-1.6}^{+3.2}\,10^{-13}$cm$^3$.s$^{-1}$. At
$636\,$G, we have initially $N_3\simeq 0.3 N_1$. These fast losses
indicate that very likely some evaporation is involved in the limited
lifetime and that the system has not reached thermal
equilibrium. Therefore, the 2-body loss rate cannot be safely estimated.

Let us now evaluate the phase space density of the trapped molecules. In
Fig.\,\ref{fig:figure2}, at $689\,$G, there are $1.8\,10^4$ molecules,
confined with $3.3\,10^4$ atoms. The measured atom temperature is
$T_{at}=6.7\,\mu$K. Assuming thermal equilibrium between atoms and
molecules, $T_{at}=T_{m}$ for magnetic fields above $\simeq 650\,$G, we
obtain the peak density of molecules $n_{m}\simeq
4\,10^{13}\,$cm$^{-3}$. Then the critical temperature for molecule
condensation is reduced due to the interactions and is $3.5\,\mu$K
giving $T/T_C\simeq 2\,$. Since $n_m a_m^3\simeq 5\,10^{-3}\ll 1$ with
$a_m=0.6\,a=47\,$nm (\cite{Petrovprivate}), the molecular gas is in the
dilute regime and the mean distance between molecules is larger than the
typical size of a molecule $a$. In fact, for all data between $675\,$G
and $750\,$G, the phase space density of the molecules is not far from
the condensation point. Since, the lifetime of the molecules, $500\,$ms,
is long in this region when compared to the molecule-molecule elastic
collision time $1/(n_m\,8\pi a_m^2 v_m)\simeq 3\,\mu$s, (for $a_m=47\,$nm), it
should be possible to evaporate the molecules further to reach the
Bose-Einstein condensation threshold.

In summary, we have produced long-lived and trapped Li$_2$ dimers. The
atom-molecule conversion efficiency can approach $1$ when the Fermi
quantum degeneracy is strong. The lifetime of the trapped molecules
strongly depends on the scattering length. Current research concentrates
on evaporation towards molecular BEC. Prospects for producing
superfluid Fermi mixtures and for investigating the transition between
molecular condensates and superfluid Fermi gases are promising
\cite{Petrovprivate,Ohashi02,Milstein02,Carr03}.

We are grateful to D.\,S.\,Petrov, S.\,Stringari, Y\,.Castin,
R.\,Combescot and J.\,Dalibard for useful discussions. This work was
supported by CNRS, Coll\`ege de France, and R\'egion Ile de
France. S.\,Kokkelmans acknowledges a Marie Curie grant from the
E.U. under contract number MCFI-2002-00968.  Laboratoire Kastler Brossel
is {\it Unit\'e de recherche de l'Ecole Normale Sup\'erieure et de
l'Universit\'e Pierre et Marie Curie, associ\'ee au CNRS}.

\end{document}